\documentclass[12pt,a4paper]{article}
\usepackage{amsmath,amsthm,amssymb}
\begin{document}
\title{On semifast Fourier transform algorithms}
\date{}

\author{Sergei V. Fedorenko \\
Information systems security department \\
St.Petersburg State University of Aerospace Instrumentation \\
190000, Bolshaya Morskaia, 67, St.Petersburg, Russia}

\maketitle

\begin{abstract}
In this paper, following 
\cite{Goertzel, Blahut83, Blahut85, Blahut08, Fedorenko02, Fedorenko03, Fedorenko06} 
we consider the relations between well-known Fourier transform algorithms.
\end{abstract}

\section{Introduction}

A semifast algorithm for a computation is an algorithm 
that significantly reduces the number of multiplications compared 
with the natural form of the computation, 
but does not reduce the number of additions. 
A semifast Fourier transform in $GF(q)$ is a computational procedure 
for computing the $n$-point Fourier transform in $GF(q)$ that
uses about $n \log n$ multiplications in $GF(q)$ and about 
$n^2$ additions in $GF(q)$ \cite{Blahut08}.

The $n$-point Fourier transform over $GF(2^m)$ is defined by 

$$F=Wf,$$
$$
\begin{pmatrix}
F_0\\ 
F_1\\
F_2\\
\cdots \\
F_{n-1}\\
\end{pmatrix}
= 
\begin{bmatrix}
 \alpha^0           & \alpha^0           & \alpha^0           & \cdots & \alpha^0               \\ 
 \alpha^0           & \alpha^1           & \alpha^2           & \cdots & \alpha^{n-1}           \\
 {(\alpha^0)}^2     & {(\alpha^1)}^2     & {(\alpha^2)}^2     & \cdots & {(\alpha^{n-1})}^2     \\
  \cdots            & \cdots             &  \cdots            & \cdots &  \cdots                \\
 {(\alpha^0)}^{n-1} & {(\alpha^1)}^{n-1} & {(\alpha^2)}^{n-1} & \cdots & {(\alpha^{n-1})}^{n-1} \\
\end{bmatrix}
\begin{pmatrix}
f_0\\ 
f_1\\
f_2\\
\cdots \\
f_{n-1}\\ 
\end{pmatrix},
$$
where 
$W = (\alpha^{ij})$, \, $i,j \in [0,n-1]$, is a Vandermonde matrix, \\
an element $\alpha \in GF(2^m)$ of order $\hbox{ord}(\alpha) = n \mid 2^m-1$
is the Fourier transform kernel.

Consider a few well-known Fourier transform algorithms 
\cite{Goertzel, Blahut83, Blahut85, Blahut08, Fedorenko02, Fedorenko03, Fedorenko06} 
on examples.

The finite field $GF(2^3)$ is defined by an element $\alpha$,
which is a root of the primitive polynomial $x^3 + x + 1$. 
Cyclotomic cosets modulo 7 over $GF(2)$ are
$(0), (1,2,4), (3,6,5)$. 
The binary conjugacy classes of $GF(2^3)$ are 
$(\alpha^0), (\alpha^1, \alpha^2, \alpha^4), 
(\alpha^3, \alpha^6, \alpha^5)$.
The standard basis of $GF(2^3)$ is 
$\left(\alpha^0, \alpha^1, \alpha^2 \right)$. 
The normal basis of $GF(2^3)$ is 
$\left(\beta^1, \beta^2, \beta^4 \right)$, where $\beta = \alpha^3$. 
Another normal basis of $GF(2^3)$ is the cyclic shift of the previous normal basis: 
$\left(\gamma^1, \gamma^2, \gamma^4 \right)$, where $\gamma = \alpha^6$. 

Consider the Fourier transform of length $n = 7$ over the field $GF(2^3)$. 
Let us take the primitive element $\alpha$ as the kernel
of the Fourier transform. 
The Fourier transform of a polynomial 
$f(x)=\displaystyle\sum_{i=0}^6 f_i x^i$ 
consists of components 
$F_i=f(\alpha^i), \quad i \in [0,6]$.

Thus,
$$
\begin{pmatrix}
F_0\\
F_1\\
F_2\\
F_3\\
F_4\\
F_5\\
F_6\\
\end{pmatrix}=
\begin{bmatrix}
 \alpha^0 & \alpha^0 & \alpha^0 & \alpha^0 & \alpha^0 & \alpha^0 & \alpha^0 \\
 \alpha^0 & \alpha^1 & \alpha^2 & \alpha^3 & \alpha^4 & \alpha^5 & \alpha^6 \\
 \alpha^0 & \alpha^2 & \alpha^4 & \alpha^6 & \alpha^1 & \alpha^3 & \alpha^5 \\
 \alpha^0 & \alpha^3 & \alpha^6 & \alpha^2 & \alpha^5 & \alpha^1 & \alpha^4 \\
 \alpha^0 & \alpha^4 & \alpha^1 & \alpha^5 & \alpha^2 & \alpha^6 & \alpha^3 \\
 \alpha^0 & \alpha^5 & \alpha^3 & \alpha^1 & \alpha^6 & \alpha^4 & \alpha^2 \\
 \alpha^0 & \alpha^6 & \alpha^5 & \alpha^4 & \alpha^3 & \alpha^2 & \alpha^1 \\
\end{bmatrix}
\begin{pmatrix}
f_0\\
f_1\\
f_2\\
f_3\\
f_4\\
f_5\\
f_6\\
\end{pmatrix}.
$$

\section{Goertzel's algorithm (1958)}
\subsection*{\hfil Blahut's modification for finite fields (1983) \hfil}

The first step of the Goertzel algorithm is a long division 
$f(x)$ by every minimal polynomial:

$$
\begin{tabular}{rcrclc} 
$f(x)$ &=& $(x+1) q_0(x)$       &+&$r_{00}$ &                                       \\
$f(x)$ &=& $(x^3+x+1) q_1(x)$   &+&$r_1(x)$,&$r_1(x) = r_{21}x^2 + r_{11}x + r_{01}$\\
$f(x)$ &=& $(x^3+x^2+1) q_2(x)$ &+&$r_2(x)$,&$r_2(x) = r_{22}x^2 + r_{12}x + r_{02}$\\
\end{tabular},
$$
where
$$
\begin{tabular}{ll} 
$r_{00}$ &= $f_0 + f_1 + f_2 + f_3 + f_4 + f_5 + f_6$ \\
$r_{01}$ &= $f_0 + f_3 + f_5+ f_6$ \\
$r_{11}$ &= $f_1 + f_3 + f_4+ f_5$ \\
$r_{21}$ &= $f_2 + f_4 + f_5+ f_6$ \\
$r_{02}$ &= $f_0 + f_3 + f_4+ f_5$ \\
$r_{12}$ &= $f_1 + f_4 + f_5+ f_6$ \\
$r_{22}$ &= $f_2 + f_3 + f_4+ f_6$.
\end{tabular}
$$

The second step of the Goertzel algorithm is 
an evaluated remainder in every element of the finite field:

$$
\begin{tabular}{cclc} 
$F_0$ &= $f(\alpha^0)$ &= $r_{00}$        &                                             \\
$F_1$ &= $f(\alpha^1)$ &= $r_1(\alpha^1)$ &= $r_{21}\alpha^2 + r_{11}\alpha^1 + r_{01}$ \\
$F_2$ &= $f(\alpha^2)$ &= $r_1(\alpha^2)$ &= $r_{21}\alpha^4 + r_{11}\alpha^2 + r_{01}$ \\
$F_4$ &= $f(\alpha^4)$ &= $r_1(\alpha^4)$ &= $r_{21}\alpha^1 + r_{11}\alpha^4 + r_{01}$ \\
$F_3$ &= $f(\alpha^3)$ &= $r_2(\alpha^3)$ &= $r_{22}\alpha^6 + r_{12}\alpha^3 + r_{02}$ \\
$F_6$ &= $f(\alpha^6)$ &= $r_2(\alpha^6)$ &= $r_{22}\alpha^5 + r_{12}\alpha^6 + r_{02}$ \\
$F_5$ &= $f(\alpha^5)$ &= $r_2(\alpha^5)$ &= $r_{22}\alpha^3 + r_{12}\alpha^5 + r_{02}$ \\
\end{tabular}
$$
or
$$
\begin{pmatrix}
F_1\\
F_2\\
F_4\\
\end{pmatrix}
=
\begin{bmatrix}
\begin{tabular}{ccc} 
 1 & $\alpha^1$ & $\alpha^2$ \\
 1 & $\alpha^2$ & $\alpha^4$ \\ 
 1 & $\alpha^4$ & $\alpha^1$ \\ 
\end{tabular}
\end{bmatrix}
\begin{pmatrix}
r_{01}\\
r_{11}\\
r_{21}\\
\end{pmatrix}
, \qquad
\begin{pmatrix}
F_3\\
F_6\\
F_5\\
\end{pmatrix}
=
\begin{bmatrix}
\begin{tabular}{ccc} 
 1 & $\alpha^3$ & $\alpha^6$ \\
 1 & $\alpha^6$ & $\alpha^5$ \\ 
 1 & $\alpha^5$ & $\alpha^3$ \\ 
\end{tabular}
\end{bmatrix}
\begin{pmatrix}
r_{02}\\
r_{12}\\
r_{22}\\
\end{pmatrix}.
$$

Thus,
$$
\begin{pmatrix}
F_0\\ \hline
F_1\\
F_2\\
F_4\\ \hline
F_3\\
F_6\\
F_5\\
\end{pmatrix}=
\begin{bmatrix}
\begin{tabular}{c|ccc|ccc} 
 $\alpha^0$ &            &            &            &            &            &           \\ \hline
            & $\alpha^0$ & $\alpha^1$ & $\alpha^2$ &            &            &           \\
            & $\alpha^0$ & $\alpha^2$ & $\alpha^4$ &            &            &           \\ 
            & $\alpha^0$ & $\alpha^4$ & $\alpha^1$ &            &            &           \\ \hline
            &            &            &            & $\alpha^0$ & $\alpha^3$ & $\alpha^6$\\
            &            &            &            & $\alpha^0$ & $\alpha^6$ & $\alpha^5$\\ 
            &            &            &            & $\alpha^0$ & $\alpha^5$ & $\alpha^3$\\ 
\end{tabular}
\end{bmatrix}
\begin{bmatrix}
1&1&1&1&1&1&1\\ \hline
1&0&0&1&0&1&1\\
0&1&0&1&1&1&0\\
0&0&1&0&1&1&1\\ \hline
1&0&0&1&1&1&0\\ 
0&1&0&0&1&1&1\\
0&0&1&1&1&0&1\\ 
\end{bmatrix}
\begin{pmatrix}
f_0\\ 
f_1\\
f_2\\
f_3\\ 
f_4\\
f_5\\
f_6\\ 
\end{pmatrix}.
$$

\section{Blahut's algorithm (2008)}

We have 

$$
\begin{pmatrix}
F_0\\
F_1\\
F_2\\
F_3\\
F_4\\
F_5\\
F_6\\
\end{pmatrix}=
\begin{bmatrix}
 \alpha^0 & \alpha^0 & \alpha^0 & \alpha^0 & \alpha^0 & \alpha^0 & \alpha^0 \\
 \alpha^0 & \alpha^1 & \alpha^2 & \alpha^3 & \alpha^4 & \alpha^5 & \alpha^6 \\
 \alpha^0 & \alpha^2 & \alpha^4 & \alpha^6 & \alpha^1 & \alpha^3 & \alpha^5 \\
 \alpha^0 & \alpha^3 & \alpha^6 & \alpha^2 & \alpha^5 & \alpha^1 & \alpha^4 \\
 \alpha^0 & \alpha^4 & \alpha^1 & \alpha^5 & \alpha^2 & \alpha^6 & \alpha^3 \\
 \alpha^0 & \alpha^5 & \alpha^3 & \alpha^1 & \alpha^6 & \alpha^4 & \alpha^2 \\
 \alpha^0 & \alpha^6 & \alpha^5 & \alpha^4 & \alpha^3 & \alpha^2 & \alpha^1 \\
\end{bmatrix}
\left[
\begin{pmatrix}
f_0\\
0  \\
0  \\
0  \\
0  \\
0  \\
0  \\
\end{pmatrix}
+
\begin{pmatrix}
0  \\
f_1\\
f_2\\
0  \\
f_4\\
0  \\
0  \\
\end{pmatrix}
+
\begin{pmatrix}
0  \\
0  \\
0  \\
f_3\\
0  \\
f_5\\
f_6\\
\end{pmatrix}
\right]
=
$$

$$
=
\begin{bmatrix}
1  \\
1  \\
1  \\
1  \\
1  \\
1  \\
1  \\
\end{bmatrix}
\begin{pmatrix}
f_0
\end{pmatrix}
+
\begin{bmatrix}
1&0&0\\
0&1&0\\
0&0&1\\
1&1&0\\
0&1&1\\
1&1&1\\
1&0&1\\  
\end{bmatrix}
\begin{bmatrix}
\begin{tabular}{ccc} 
 $\alpha^0$ & $\alpha^0$ & $\alpha^0$ \\
 $\alpha^1$ & $\alpha^2$ & $\alpha^4$ \\ 
 $\alpha^2$ & $\alpha^4$ & $\alpha^1$ \\ 
\end{tabular}
\end{bmatrix}
\begin{pmatrix}
f_1\\
f_2\\
f_4\\
\end{pmatrix}
+
$$

$$
+
\begin{bmatrix}
1&0&0\\
0&1&0\\
0&0&1\\
1&0&1\\
1&1&1\\
1&1&0\\
0&1&1\\  
\end{bmatrix}
\begin{bmatrix}
\begin{tabular}{ccc} 
 $\alpha^0$ & $\alpha^0$ & $\alpha^0$ \\
 $\alpha^3$ & $\alpha^6$ & $\alpha^5$ \\ 
 $\alpha^6$ & $\alpha^5$ & $\alpha^3$ \\ 
\end{tabular}
\end{bmatrix}
\begin{pmatrix}
f_3\\
f_6\\
f_5\\
\end{pmatrix}.
$$

\section{Fedorenko--Trifonov's algorithm (2002)}

A polynomial $L(x)=\displaystyle\sum_{i} l_i x^{2^i}$, \,
$l_i \in GF(2^m)$, is called a linearized polynomial.
Any polynomial can be decomposed into a sum of linearized polynomials
and a free term.
A polynomial $f(x)=\displaystyle\sum_{i=0}^6 f_i x^i$
can be represented as

$$
\begin{tabular}{rcccccc} 
$f(x)$  &=&$L_0(x^0)$&+&$L_1(x^1)$&+&$L_2(x^3)$\\
$L_0(y)$&=&$f_0$     & &          & &$        $\\
$L_1(y)$&=&$f_1y$    &+&$f_2y^2$  &+&$f_4y^4  $\\
$L_2(y)$&=&$f_3y$    &+&$f_6y^2$  &+&$f_5y^4  $.
\end{tabular}
$$

We have 

\begin{eqnarray*}
f(\alpha^0)=&L_0(\alpha^0)+L_1(\alpha^0)+L_2(\alpha^0)\phantom{=}	&\\
f(\alpha^1)=&L_0(\alpha^0)+L_1(\alpha^1)+L_2(\alpha^3)=&L_0(1)+L_1(\alpha)+L_2(1)+L_2(\alpha) \\
f(\alpha^2)=&L_0(\alpha^0)+L_1(\alpha^2)+L_2(\alpha^6)=&L_0(1)+L_1(\alpha^2)+L_2(1)+L_2(\alpha^2) \\
f(\alpha^3)=&L_0(\alpha^0)+L_1(\alpha^3)+L_2(\alpha^2)=&L_0(1)+L_1(1)+L_1(\alpha)+L_2(\alpha^2) \\
f(\alpha^4)=&L_0(\alpha^0)+L_1(\alpha^4)+L_2(\alpha^5)=&L_0(1)+L_1(\alpha)+L_1(\alpha^2)+ \\
            &                                          &L_2(1)+L_2(\alpha)+L_2(\alpha^2) \\
f(\alpha^5)=&L_0(\alpha^0)+L_1(\alpha^5)+L_2(\alpha^1)=&L_0(1)+L_1(1)+L_1(\alpha)+L_1(\alpha^2)+L_2(\alpha) \\
f(\alpha^6)=&L_0(\alpha^0)+L_1(\alpha^6)+L_2(\alpha^4)=&L_0(1)+L_1(1)+L_1(\alpha^2)+L_2(\alpha)+L_2(\alpha^2).
\end{eqnarray*}

Using $F_i=f(\alpha^i)$, these equations can be represented in a matrix form as

$$
\begin{pmatrix}
F_0\\
F_1\\
F_2\\
F_3\\
F_4\\
F_5\\
F_6\\
\end{pmatrix}=
\begin{bmatrix}
\begin{tabular}{c|ccc|ccc} 
1&1&0&0&1&0&0\\
1&0&1&0&1&1&0\\
1&0&0&1&1&0&1\\
1&1&1&0&0&0&1\\
1&0&1&1&1&1&1\\
1&1&1&1&0&1&0\\
1&1&0&1&0&1&1\\
\end{tabular}
\end{bmatrix}
\begin{bmatrix}
\begin{tabular}{c|ccc|ccc} 
 $\alpha^0$ &            &            &            &            &            &           \\ \hline
            & $\alpha^0$ & $\alpha^0$ & $\alpha^0$ &            &            &           \\
            & $\alpha^1$ & $\alpha^2$ & $\alpha^4$ &            &            &           \\ 
            & $\alpha^2$ & $\alpha^4$ & $\alpha^1$ &            &            &           \\ \hline
            &            &            &            & $\alpha^0$ & $\alpha^0$ & $\alpha^0$\\
            &            &            &            & $\alpha^1$ & $\alpha^2$ & $\alpha^4$\\ 
            &            &            &            & $\alpha^2$ & $\alpha^4$ & $\alpha^1$\\ 
\end{tabular}
\end{bmatrix}
\begin{pmatrix}
f_0\\ \hline
f_1\\
f_2\\
f_4\\ \hline
f_3\\
f_6\\
f_5\\ 
\end{pmatrix}.
$$

\section{Trifonov--Fedorenko's algorithm (2003)}

A circulant matrix, or a circulant, is a matrix each row 
of which is obtained from the preceding 
row by a left (right) cyclic shift by one position.
If entries of a circulant are matrices, the circulant 
is referred to as a block circulant.
We call a circulant where the first row is a normal basis, 
a basis circulant.

Let $\left(\beta^1,\beta^2,\beta^4\right)$ = 
$\left(\alpha^{3},\alpha^{6},\alpha^{5}\right)$
be a normal basis for $GF(2^3)$.

Using 

$$
\begin{bmatrix}
\begin{tabular}{ccc} 
$\alpha^0$ & $\alpha^0$ & $\alpha^0$ \\
$\alpha^1$ & $\alpha^2$ & $\alpha^4$ \\ 
$\alpha^2$ & $\alpha^4$ & $\alpha^1$ \\ 
\end{tabular}
\end{bmatrix}
=
\begin{bmatrix}
\begin{tabular}{ccc} 
1&1&1 \\
0&1&1 \\
1&0&1 \\ 
\end{tabular}
\end{bmatrix}
\begin{bmatrix}
\begin{tabular}{ccc} 
$\beta^1$ & $\beta^2$ & $\beta^4$ \\
$\beta^2$ & $\beta^4$ & $\beta^1$ \\
$\beta^4$ & $\beta^1$ & $\beta^2$ \\ 
\end{tabular}
\end{bmatrix},
$$

we get

\begin{equation*}
\begin{pmatrix}
F_0\\
F_1\\
F_2\\
F_3\\
F_4\\
F_5\\
F_6
\end{pmatrix}=
\begin{bmatrix}
\begin{tabular}{c|ccc|ccc} 
1&1&1&1&1&1&1\\
1&0&1&1&1&0&0\\
1&1&0&1&0&1&0\\
1&1&0&0&1&0&1\\
1&1&1&0&0&0&1\\
1&0&0&1&0&1&1\\
1&0&1&0&1&1&0
\end{tabular} 
\end{bmatrix}
\begin{bmatrix}
\begin{tabular}{c|ccc|ccc} 
1&          &          &         &         &         &         \\ \hline
 &$\beta^1$ &$\beta^2$ &$\beta^4$&         &         &         \\
 &$\beta^2$ &$\beta^4$ &$\beta^1$&         &         &         \\
 &$\beta^4$ &$\beta^1$ &$\beta^2$&         &         &         \\ \hline
 &          &          &         &$\beta^1$&$\beta^2$&$\beta^4$\\
 &          &          &         &$\beta^2$&$\beta^4$&$\beta^1$\\
 &          &          &         &$\beta^4$&$\beta^1$&$\beta^2$\\
\end{tabular}
\end{bmatrix}
\begin{pmatrix}
f_0\\ \hline
f_1\\
f_2\\
f_4\\ \hline
f_3\\
f_6\\
f_5\\
\end{pmatrix}.
\end{equation*}

\section{Fedorenko's algorithm (1) (2006)}

Combining

$$
\begin{bmatrix}
 \alpha^3 & \alpha^6 & \alpha^5 \\
 \alpha^6 & \alpha^5 & \alpha^3 \\
 \alpha^5 & \alpha^3 & \alpha^6 \\
\end{bmatrix}
=
\begin{bmatrix}
 \beta^1 & \beta^2 & \beta^4 \\
 \beta^2 & \beta^4 & \beta^1 \\
 \beta^4 & \beta^1 & \beta^2 \\
\end{bmatrix}
, \qquad
\begin{bmatrix}
 \alpha^1 & \alpha^2 & \alpha^4 \\
 \alpha^2 & \alpha^4 & \alpha^1 \\
 \alpha^4 & \alpha^1 & \alpha^2 \\
\end{bmatrix}
=
\begin{bmatrix}
 0 & 1 & 1 \\
 1 & 0 & 1 \\
 1 & 1 & 0 \\
\end{bmatrix}
\begin{bmatrix}
 \beta^1 & \beta^2 & \beta^4 \\
 \beta^2 & \beta^4 & \beta^1 \\
 \beta^4 & \beta^1 & \beta^2 \\
\end{bmatrix},
$$

$$
\begin{bmatrix}
 \alpha^2 & \alpha^4 & \alpha^1 \\ 
 \alpha^4 & \alpha^1 & \alpha^2 \\ 
 \alpha^1 & \alpha^2 & \alpha^4 \\ 
\end{bmatrix}
=
\begin{bmatrix}
 1 & 0 & 1 \\
 1 & 1 & 0 \\
 0 & 1 & 1 \\
\end{bmatrix}
\begin{bmatrix}
 \beta^1 & \beta^2 & \beta^4 \\
 \beta^2 & \beta^4 & \beta^1 \\
 \beta^4 & \beta^1 & \beta^2 \\
\end{bmatrix},
$$

we obtain the equivalent Fourier transform

\begin{center}
$
\begin{pmatrix}
F_0\\ \hline
F_1\\
F_2\\
F_4\\ \hline
F_3\\
F_6\\
F_5\\ 
\end{pmatrix}
$ =
$
\begin{bmatrix}
\begin{tabular}{c|ccc|ccc} 
$\alpha^0$ & $\alpha^0$ & $\alpha^0$ & $\alpha^0$ & $\alpha^0$ & $\alpha^0$ & $\alpha^0$ \\ \hline
$\alpha^0$ & $\alpha^1$ & $\alpha^2$ & $\alpha^4$ & $\alpha^3$ & $\alpha^6$ & $\alpha^5$ \\
$\alpha^0$ & $\alpha^2$ & $\alpha^4$ & $\alpha^1$ & $\alpha^6$ & $\alpha^5$ & $\alpha^3$ \\
$\alpha^0$ & $\alpha^4$ & $\alpha^1$ & $\alpha^2$ & $\alpha^5$ & $\alpha^3$ & $\alpha^6$ \\ \hline
$\alpha^0$ & $\alpha^3$ & $\alpha^6$ & $\alpha^5$ & $\alpha^2$ & $\alpha^4$ & $\alpha^1$ \\
$\alpha^0$ & $\alpha^6$ & $\alpha^5$ & $\alpha^3$ & $\alpha^4$ & $\alpha^1$ & $\alpha^2$ \\
$\alpha^0$ & $\alpha^5$ & $\alpha^3$ & $\alpha^6$ & $\alpha^1$ & $\alpha^2$ & $\alpha^4$ \\ 
\end{tabular}
\end{bmatrix}
$
$
\begin{pmatrix}
f_0\\ \hline
f_1\\
f_2\\
f_4\\ \hline
f_3\\
f_6\\
f_5\\ 
\end{pmatrix}
$ =
\end{center}

\begin{center}
=
$
\begin{bmatrix} 
\begin{tabular}{c|ccc|ccc} 
 1 & 1 & 1 & 1 & 1 & 1 & 1 \\ \hline
 1 & 0 & 1 & 1 & 1 & 0 & 0 \\
 1 & 1 & 0 & 1 & 0 & 1 & 0 \\
 1 & 1 & 1 & 0 & 0 & 0 & 1 \\ \hline
 1 & 1 & 0 & 0 & 1 & 0 & 1 \\
 1 & 0 & 1 & 0 & 1 & 1 & 0 \\
 1 & 0 & 0 & 1 & 0 & 1 & 1 \\ 
\end{tabular}
\end{bmatrix} 
$
$
\begin{bmatrix}
\begin{tabular}{c|ccc|ccc} 
1&          &          &          &          &          &          \\ \hline
 &$\beta^1$ &$\beta^2$ &$\beta^4$ &          &          &          \\
 &$\beta^2$ &$\beta^4$ &$\beta^1$ &          &          &          \\
 &$\beta^4$ &$\beta^1$ &$\beta^2$ &          &          &          \\ \hline
 &          &        &            &$\beta^1$ &$\beta^2$ &$\beta^4$ \\
 &          &        &            &$\beta^2$ &$\beta^4$ &$\beta^1$ \\
 &          &        &            &$\beta^4$ &$\beta^1$ &$\beta^2$ \\
\end{tabular}
\end{bmatrix}
$
$
\begin{pmatrix}
f_0\\ \hline
f_1\\
f_2\\
f_4\\ \hline
f_3\\
f_6\\
f_5\\ 
\end{pmatrix}
$.
\end{center}

The equivalent Fourier transform $F_e = W_e f_e$ has the following structure:

\begin{equation}
\label{1}
F_e = W_e f_e = A_e D_e f_e =
\begin{bmatrix}
 A_{11} & \cdots & A_{1l} \\
 \cdots & \cdots & \cdots \\
 A_{l1} & \cdots & A_{ll} \\
\end{bmatrix}
\begin{bmatrix}
C_1   & 0    &\ldots & 0 \\
0     & C_2  &\ldots & 0 \\
\ldots&\ldots&\ddots & \ldots\\
0     & 0    &\ldots & C_l \\
\end{bmatrix}
f_e,
\end{equation}
where 
$A_e$ is a binary matrix, \\
$A_e$ consists of binary circulants $A_{ij}$, \\
$D_e$ is a block diagonal matrix, \\
$D_e$ consists of basis circulant matrices $C_i$, \\
$l$ is the number of cyclotomic cosets modulo $n$ over $GF(2)$.

\section{Fedorenko's algorithm (2) (2006)}

Let $\left(\gamma^1,\gamma^2,\gamma^4\right)$ = 
$\left(\alpha^{6},\alpha^{5},\alpha^{3}\right)$
be a normal basis for $GF(2^3)$.

Combining

$$
\begin{bmatrix}
 \alpha^6 & \alpha^5 & \alpha^3 \\
 \alpha^5 & \alpha^3 & \alpha^6 \\
 \alpha^3 & \alpha^6 & \alpha^5 \\
\end{bmatrix}
=
\begin{bmatrix}
 \gamma^1 & \gamma^2 & \gamma^4 \\
 \gamma^2 & \gamma^4 & \gamma^1 \\
 \gamma^4 & \gamma^1 & \gamma^2 \\
\end{bmatrix}
, \qquad
\begin{bmatrix}
 \alpha^1 & \alpha^2 & \alpha^4 \\
 \alpha^2 & \alpha^4 & \alpha^1 \\
 \alpha^4 & \alpha^1 & \alpha^2 \\
\end{bmatrix}
=
\begin{bmatrix}
 1 & 1 & 0 \\
 0 & 1 & 1 \\
 1 & 0 & 1 \\
\end{bmatrix}
\begin{bmatrix}
 \gamma^1 & \gamma^2 & \gamma^4 \\
 \gamma^2 & \gamma^4 & \gamma^1 \\
 \gamma^4 & \gamma^1 & \gamma^2 \\
\end{bmatrix},
$$

we obtain

\begin{center}
$
\begin{pmatrix}
F_0\\ \hline
F_1\\
F_2\\
F_4\\ \hline
F_6\\
F_5\\
F_3\\ 
\end{pmatrix}
$ =
$
\begin{bmatrix}
\begin{tabular}{c|ccc|ccc} 
$\alpha^0$ & $\alpha^0$ & $\alpha^0$ & $\alpha^0$ & $\alpha^0$ & $\alpha^0$ & $\alpha^0$ \\ \hline
$\alpha^0$ & $\alpha^1$ & $\alpha^2$ & $\alpha^4$ & $\alpha^6$ & $\alpha^5$ & $\alpha^3$ \\
$\alpha^0$ & $\alpha^2$ & $\alpha^4$ & $\alpha^1$ & $\alpha^5$ & $\alpha^3$ & $\alpha^6$ \\
$\alpha^0$ & $\alpha^4$ & $\alpha^1$ & $\alpha^2$ & $\alpha^3$ & $\alpha^6$ & $\alpha^5$ \\ \hline
$\alpha^0$ & $\alpha^6$ & $\alpha^5$ & $\alpha^3$ & $\alpha^1$ & $\alpha^2$ & $\alpha^4$ \\
$\alpha^0$ & $\alpha^5$ & $\alpha^3$ & $\alpha^6$ & $\alpha^2$ & $\alpha^4$ & $\alpha^1$ \\
$\alpha^0$ & $\alpha^3$ & $\alpha^6$ & $\alpha^5$ & $\alpha^4$ & $\alpha^1$ & $\alpha^2$ \\ 
\end{tabular}
\end{bmatrix}
$
$
\begin{pmatrix}
f_0\\ \hline
f_1\\
f_2\\
f_4\\ \hline
f_6\\
f_5\\
f_3\\ 
\end{pmatrix}
$ =
\end{center}

\begin{center}
=
$
\begin{bmatrix} 
\begin{tabular}{c|ccc|ccc} 
 1 & 1 & 1 & 1 & 1 & 1 & 1 \\ \hline
 1 & 1 & 1 & 0 & 1 & 0 & 0 \\
 1 & 0 & 1 & 1 & 0 & 1 & 0 \\
 1 & 1 & 0 & 1 & 0 & 0 & 1 \\ \hline
 1 & 1 & 0 & 0 & 1 & 1 & 0 \\
 1 & 0 & 1 & 0 & 0 & 1 & 1 \\
 1 & 0 & 0 & 1 & 1 & 0 & 1 \\ 
\end{tabular}
\end{bmatrix} 
$
$
\begin{bmatrix}
\begin{tabular}{c|ccc|ccc} 
1&          &          &          &          &          &          \\ \hline
 &$\gamma^1$&$\gamma^2$&$\gamma^4$&          &          &          \\
 &$\gamma^2$&$\gamma^4$&$\gamma^1$&          &          &          \\
 &$\gamma^4$&$\gamma^1$&$\gamma^2$&          &          &          \\ \hline
 &          &        &            &$\gamma^1$&$\gamma^2$&$\gamma^4$\\
 &          &        &            &$\gamma^2$&$\gamma^4$&$\gamma^1$\\
 &          &        &            &$\gamma^4$&$\gamma^1$&$\gamma^2$\\
\end{tabular}
\end{bmatrix}
$
$
\begin{pmatrix}
f_0\\ \hline
f_1\\
f_2\\
f_4\\ \hline
f_6\\
f_5\\
f_3\\ 
\end{pmatrix}
$.
\end{center}

Note that the binary matrix $A_e$ for this algorithm in (\ref{1}) 
consists of binary circulants and block circulants.

\section{Complexity}

The Fourier transform algorithms \cite{Fedorenko02, Fedorenko03, Fedorenko06} 
of length $n=2^m-1$ over $GF(2^m)$ take two stages:

\begin{enumerate}
\item The first stage is calculation of $l$ $m$-point cyclic convolutions;
\item The second stage is multiplying the binary matrix $A_e$ by the vector 
$D_e f_e$.
\end{enumerate}

The complexity of the first stage is about $n \log n$ 
multiplications and additions over elements of $GF(2^m)$.
The complexity of the second stage is $N_{add} < 2 n^2 / \log n$
additions over elements of $GF(2^m)$ \cite{AHU}.

\end{document}